\documentclass[%
 aip,
 amsmath,amssymb,
 reprint,%
]{revtex4-2}

\usepackage{graphicx}
\usepackage{dcolumn}
\usepackage{bm}
\usepackage[utf8]{inputenc}
\usepackage[T1]{fontenc}
\usepackage{mathptmx}
\usepackage{etoolbox}
\usepackage{xcolor}
\usepackage{hyperref}

\makeatletter
\def\@email#1#2{%
 \endgroup
 \patchcmd{\titleblock@produce}
  {\frontmatter@RRAPformat}
  {\frontmatter@RRAPformat{\produce@RRAP{*#1\href{mailto:#2}{#2}}}\frontmatter@RRAPformat}
  {}{}
}
\makeatother
\begin{document}

\preprint{AIP/123-QED}

\title{Dynamics of self-propelled tracer particles inside a polymer network}
\author{Praveen Kumar}
 \affiliation{Department of Chemistry, Indian Institute of Technology Bombay, Mumbai, Maharashtra -  400076, India}
\author{Rajarshi Chakrabarti$^\ast$}
\email{rajarshi@chem.iitb.ac.in}
\affiliation{Department of Chemistry, Indian Institute of Technology Bombay, Mumbai, Maharashtra -  400076, India}

\begin{abstract}
\noindent Transport of tracer particles through mesh-like environments such as biological hydrogels and polymer matrices is ubiquitous in nature. These tracers could be passive, such as colloids or active (self-propelled), such as synthetic nanomotors or bacteria. Computer simulations in principle should be extremely useful in exploring the mechanism of active (self-propelled) transport of tracer particles through the mesh-like environments. Therefore, we construct a polymer network on a diamond lattice and use computer simulations to investigate the dynamics of spherical self-propelled particles inside the network. Our main objective is to elucidate the effect of the self-propulsion on the dynamics of the tracer particle as a function of tracer size and stiffness of the polymer network. We compute the time-averaged mean-squared displacement (MSD) and the van-Hove correlations of the tracer. On one hand, in the case of the bigger sticky particle, caging caused by the network particles wins over the escape assisted by the self-propulsion. This results intermediate-time subdiffusion. On the other hand, smaller tracers or tracers with high self-propulsion velocities can easily escape from the cages and show intermediate-time superdiffusion. Stiffer the network, slower the dynamics of the tracer, and the bigger tracers  exhibit longer lived intermediate time superdiffusion, as the persistence time scales as $\sim \sigma^3$, where $\sigma$ is the diameter of the tracer. In intermediate time, non-Gaussianity is more pronounced for active tracers. In the long time, the dynamics of the tracer, if passive or weakly active, becomes Gaussian and diffusive, but remains flat for tracers with high self-propulsion, accounting for their seemingly unrestricted motion inside the network.
\end{abstract}

\maketitle

\section{Introduction} \label{sec:Introduction}
\noindent Understanding the dynamics of macromolecules and nanoparticles through crowded complex environments such as polymer matrix (solutions, melts, or networks), and colloidal suspensions, is of great interest to researchers, due to its implications across the scientific disciplines, including biophysics, materials science, chemistry, and medical engineering.\cite{amblard1996subdiffusion,wong2004anomalous,mitchell2021engineering,pederson2000diffusional,cherstvy2019non,yamamoto2015microscopic,kwon2014dynamics,kumar2019transport,sorichetti2021dynamics,godec2014collective} Living cells are occupied by a variety of macromolecules, namely proteins, enzymes, chromosome-filled nucleus, \textit{etc.}, which are immersed in cytoplasmic fluid. The diffusion of these biomolecules plays pivotal role in various physiochemical processes, such as protein-protein association, enzyme reactions, gene transcription, and signal transmission.\cite{wang2012disordered,berry2002monte,sigalov2010protein,cluzel2000ultrasensitive} In \textit{vitro}, several recent experimental studies have focused on the diffusion of passive agents like nanoparticles in viscoelastic medium, made of polymeric chains, mimicking mucus membrane, nuclear pore complex etc. \cite{lieleg2011biological,bansil2018biology,wagner2017rheological,wong2004anomalous,kowalczyk2011single} In general, these studies based on fluorescence correlation spectroscopy (FCS) and single-particle tracking (SPT) add to our understanding of the transport mechanism of specific molecules (drug-carriers), pathogens, proteins through biological hydrogels. The selective transport of these molecules greatly depends on their size, affinity towards the network, mesh-size and elasticity of the network \textit{etc.}\cite{lieleg2011biological,tang2009biodegradable,witten2017particle,milster2021tuning,amsden1998solute,stylianopoulos2010diffusion,kim2020tuning} \\ 

\noindent Transport process becomes even more intriguing when the tracers are driven out of equilibrium. Examples include, motor assisted macromolecules,\cite{zheng2000prestin,sundararajan2008catalytic} bacteria,\cite{sokolov2007concentration,loose2014bacterial} spermatozoa,\cite{woolley2003motility,riedel2005self} micro-tubules\cite{sumino2012large}, and active filaments\cite{mizuno2007nonequilibrium,loose2014bacterial} along with numerous artificial microswimmers such as half-coated Janus spheres,\cite{gomez2016dynamics,singh2022interaction} chiral particles,\cite{ghosh2009controlled} active colloids,\cite{howse2007self} catalytic nanomotors,\cite{wang2015fabrication,sundararajan2008catalytic} and vesicles.\cite{joseph2017chemotactic} These objects are capable of generating directed motion by drawing energy from their environments, such as either by consuming ATP,\cite{sumino2012large,mizuno2007nonequilibrium} using chemical reactions,\cite{howse2007self,guix2018self} due to concentration gradient,\cite{sokolov2007concentration,guix2018self} self-electrophoresis,\cite{wang2006bipolar} by light-induced asymmetric photodecomposition,\cite{lozano2016phototaxis,guix2018self} or due to temperature gradients \cite{jiang2010active}. Obviously these objects are out of equilibrium and some of these exhibit self-propelled motion, examples include bacteria, active colloids, Janus particles etc. \\ 

\noindent Some of the recent experimental and simulation studies focused on the transport mechanism of the self-propelled particle in viscoelastic and crowded media.\cite{patteson2016active,du2019study,yuan2019activity,gomez2016dynamics,singh2022interaction,theeyancheri2020translational,wu2021mechanisms,cao2021chain,bechinger2016active,kumar2022chemically,aragones2018diffusion,sahoo2022transport,theeyancheri2022silico} Presence of crowders make the dynamics of these self-propelled probes quite complex. Not only the translational motion but the rotational motion of these active probes get enhanced as found in experiments \cite{gomez2016dynamics,bechinger2022rod} and further supported by simulations.\cite{theeyancheri2020translational} A similar observation has been reported in a theoretical and computational investigation of self-propelled tracers in densely packed nonmotile solid particles\cite{aragones2018diffusion}. On the other hand, an experiment on \textit{E. coli} in a polymeric solution demonstrated an enhancement in cell translational diffusion and a sharp decrease in rotational diffusion.\cite{patteson2015running}  Adding to these, some studies of the self-propelled particles in crowded media have predicted non-monotonous behavior of the rotational diffusivity with the area fraction of the crowders.\cite{abaurrea2020autonomously,theeyancheri2020translational} Very recently there have been attempts to investigate the dynamics of self-propelled particles in polymer network by computer simulations.\cite{kumar2022chemically,jeon2022active} But still a comprehensive picture of active tracer dynamics over a range of activity in dense polymeric network is largely lacking. \\

\noindent In the present work, we investigate the dynamics of self-propelled tracer particles inside a dense polymer network using computer simulations. The network is constructed on a diamond lattice, where each lattice site is occupied by a monomer (bead) and each monomer is connected to the four nearest neighboring monomers by finitely extensible nonlinear elastic (FENE) springs. The self-propelled tracer particles are modeled as a spherical active Brownian particles (ABPs) (Fig.~\ref{fig:3D_Gel}). Here, the total volume occupied by the polymer network is $\sim$ 33\% (see Section \ref{sec:Simulation} for details), comparable to the crowder fractions in living cells, where 30–40\% of the total volume is occupied by biopolymers, such as proteins, nucleic acids, ribosomes, and other crowders. In our simulations, the polymer network is not frozen and after equilibration, it has fluctuating mesh sizes ranging from 0.9 to 1.8$\sigma$. Though, our gel is constructed on a lattice, but the fluctuating mesh sizes provide some extent of heterogeneity, as found in case of real polymer gels. Our mesh-like structured environment is quite different from the previously considered polymeric systems in computational studies. Earlier, people have considered network formed by connecting polymer chains or double-spring polymer network models with well-defined mesh size and network topology.\cite{milster2021tuning,cho2020tracer,cao2021chain,hu2022doublespring}  In particular, we aim at elucidating the tug of war between self-propulsion force and the binding affinity of the tracer with the network and on the process of transport. In addition, the size of the tracer and the strength of self-propulsion are expected to play important role. In the current study, we carry out Langevin dynamics simulations in the overdamped limit which is more realistic for all practical purposes. We found that the dynamics of the tracer crossover short-time subdiffusion to an intermediate-time superdiffusion. We attribute that subdiffusion is due to sticky confined motion of the tracer which eventually changes to superdiffusion in case of active tracers. There is a competition between the suppressed motion of the tracer inside a mesh and its tendency to escape from it due to self-propulsion. Interestingly, when the activity is high, the particle always undergoes superdiffusive dynamics at the intermediate time, while for weakly active tracer a short time subdiffusion emerges before it becomes superdiffusive. On the other hand, for the sticky passive tracer, the dynamics is never superdiffusive but subdiffusive before it becomes diffusive in the long time. The bigger tracers, if sticky, show stronger subdiffusion and if active also exhibits stronger superdiffusion compared to a tracer of smaller size. Our analyses of the self-part of the van-Hove correlation function of the tracers show that the the correlations increasingly become broader and non-Gaussian on increasing the magnitude of active force but approaches Gaussianity in the long time, if the activity is zero or moderate. Tracers with high self-propelled velocities tend to have even broader and flat distributions in the long time, indicating free unhindered transport through the network. \\
\noindent This paper is organized as follows. In Section 2, we present the model and simulation details. Results and discussion are presented in Section 3 and lastly, we conclude the paper in Section 4.

\section{Model and simulation details} \label{sec:Simulation} 
\begin{figure*}[ht]
    \centering
    \includegraphics[width=0.99\linewidth]{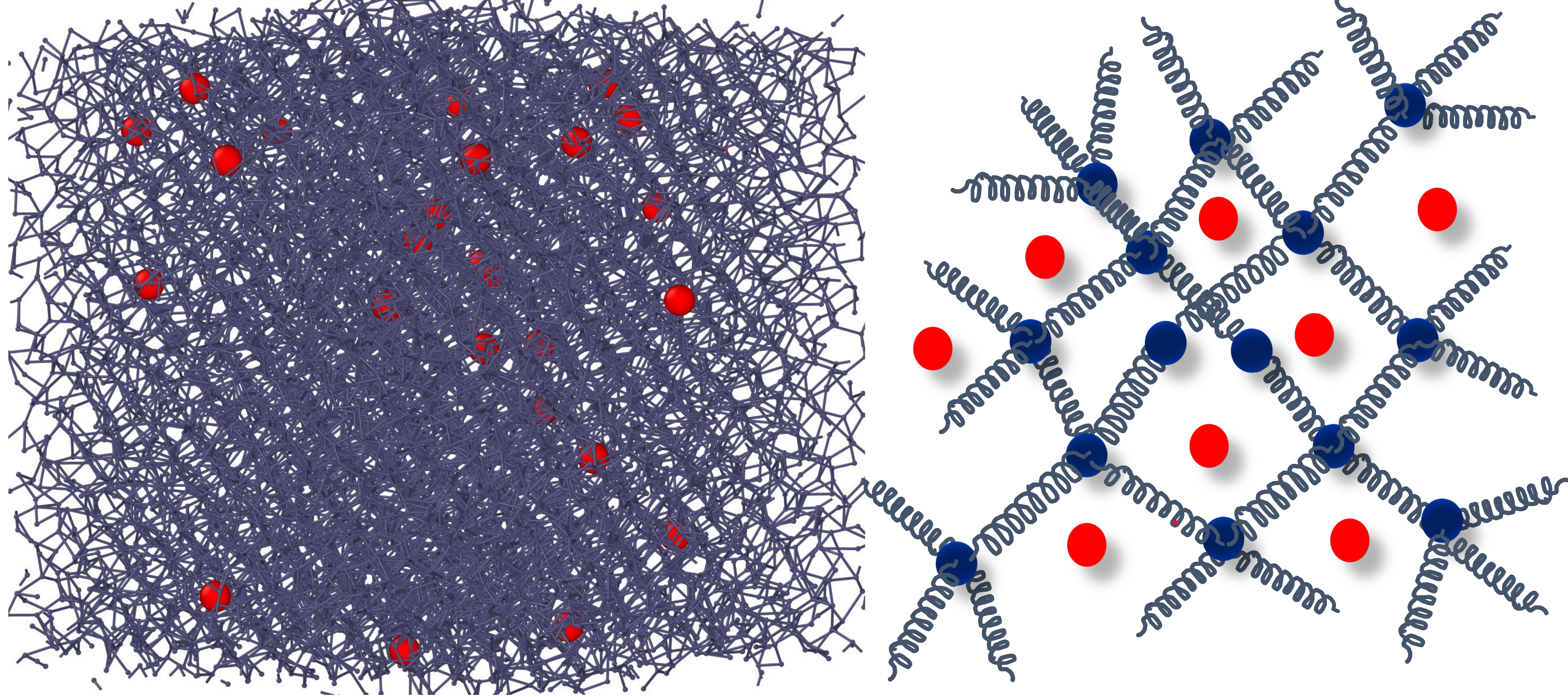}
    \caption{A snapshot (left) of the self-propelled tracer particles (red) inside the 3D polymer network (blue and springs are shown as solid lines) and 2D schematic diagram (right) of the model. This snapshot is created using Ovito (Open Visualization Tool).\cite{stukowski2009visualization}}
    \label{fig:3D_Gel}
\end{figure*}

\noindent We mimic the crowded and complex mesh-like environment by creating a 3D polymer network on the diamond lattice consists of $M$ ($M$ = 9883) monomers of size (diameter) $\sigma$, where each lattice site is occupied by a monomer and each monomer is connected to the nearest four neighboring monomers through FENE springs:\cite{kremer1990dynamics}

\begin{equation}
V_{\text{FENE}}\left(r_{ij} \right)=\begin{cases} -\frac{\text{k} r_{\text{max}}^2}{2} \log\left[1-\left( {\frac{r_{ij}}{r_{\text{max}}}}\right) ^2 \right],\hspace{5mm} \mbox{if } r_{ij} \leq r_{\text{max}}\\
=\infty, \hspace{41mm} \mbox{otherwise}
\end{cases}
\label{eq:FENE}
\end{equation}

\noindent where $r_{ij} = \vert \textbf{r}_{i} - \textbf{r}_{j}\vert$ represents the separation between two monomers i and j (with position vectors given by $\textbf{r}_{i}$ and $\textbf{r}_{j}$, respectively) of the polymer network. $r_{\text{max}} = 2.5 \sigma$ is the upper limit of $r_{ij}$ and k denotes the force constant, which accounts for the stiffness of the gel. We set a purely repulsive pairwise non-bonded interactions between the monomers of the polymer network, modeled by the Weeks–Chandler–Andersen (WCA) potential:\cite{weeks1971role}

\begin{equation}
V_{\text{WCA}}(r_{ij})=\begin{cases}4\epsilon_{ij}\left[\left(\frac{\sigma_{ij}}{r_{ij}}\right)^{12}-\left(\frac{\sigma_{ij}}{r_{ij}}\right)^{6}\right]+\epsilon_{ij}, \hspace{5mm} \mbox{if } r_{ij} \leq (2)^{1/6}\sigma_{ij}\\
=0, \hspace{50mm} \mbox{otherwise}
\end{cases}
\label{eq:WCA}
\end{equation}
\noindent for WCA $i=j$ and $\epsilon_{ii}=\epsilon$. Next, a total of $\text{N}_\text{P} = 25$ tracer particles (repulsive to each other) are randomly placed inside the polymer network to get better statistics (Fig.~\ref{fig:3D_Gel}). The system has been packed into a periodic simulation cubic box of length, $L = 25\sigma$. Hence, the total volume occupied by the polymer network is, $\dfrac{M \pi \sigma^3}{6 \times L^3} \sim 33\%$. The non-bonded attractive interactions between the monomers of the gel and tracers particles are modeled by Lennard-Jones (LJ) potential: 
 \begin{equation}
V_{\text{LJ}}^{ij}(r_{ij})=\begin{cases}4\epsilon_{ij}\left[\left(\frac{\sigma_{ij}}{r_{ij}}\right)^{12}-\left(\frac{\sigma_{ij}}{r_{ij}}\right)^{6}\right], \hspace{5mm} \mbox{if } r_{ij} \leq r_{\text{cut}}^{ij}\\
=0, \hspace{41mm} \mbox{otherwise}\\
\end{cases}
\label{eq:LJ}
\end{equation}	
\noindent here, the subscripts $i$ and $j$ represent both the monomers and the tracer particles, $r_{ij}$ and $\epsilon_{ij}$ indicate the distance and strength of the attractive interaction or binding affinity between two particles $i$ and $j$, respectively. $\sigma_{ii}$ is the diameter of the particle $i$ and $\sigma_{ij}$ is the sum of the radii of two interacting particles, $\sigma_{ij} = \frac{1}{2}(\sigma_{i}+\sigma_{j})$ and $r_{\text{cut}}^{ij} = 2.5 \sigma_{ij}$ is the cutoff radius for monomer-tracer particle pair interaction. \\

\noindent We implement the following Langevin equation to describe the motion of the $i^{th}$ particle with the identical mass $m$ and the position $r_{i}(t) $ at time t, interacting with all the other particles in the system:
	\begin{equation}
	m\frac{d^2 \textbf{r}_{i}(t)}{dt^2}=-\gamma\frac{d\textbf{r}_{i}}{dt}-\sum_{j} \nabla_{i} V(\textbf{r}_{i}-\textbf{r}_{j}) + f_{i}(t) + \textbf{F}_\text{a} \textbf{n}
	\label{eq:langevineq3}
	\end{equation} 
\noindent here, $\textbf{r}_j$ represents the position of all the particles except the $i^{th}$ particle in the system, $\gamma$ is the friction coefficient of the particle in the background implicit pure solvent, which is very high ($1.06 \times 10^{4}$) in our simulations, so that the dynamics is effectively overdamped. The total potential energy of the system can be written as $V(r) = V_{\text{FENE}} + V_{\text{LJ}} + V_{\text{WCA}}$, where $V_{\text{FENE}}$ is spring potential for the polymer gel, $V_{\text{LJ}}$ is the attractive potential, and $V_{\text{WCA}}$ corresponds to repulsive interactions. $f_{i}(t)$ is the Gaussian thermal noise with the statistical properties, 

\begin{equation}
	\left<f(t)\right>=0, \hspace{5mm}
	\left<f_{\alpha}(t^{\prime})f_{\beta}(t^{\prime\prime})\right>=6 \gamma k_B T\delta_{\alpha\beta}\delta(t^{\prime}-t^{\prime\prime})
	\label{eq:random-force}
\end{equation}
\noindent where k$_\text{B}$ is the Boltzmann constant, $\text{T}$ is the system temperature, and $\delta$ represents the Dirac delta-function, $\alpha$ and $\beta$ represent the Cartesian components.
\noindent In Eq.~\ref{eq:langevineq3}, $\textbf{F}_{a}$ is the amplitude of the active force (self-propulsion) which acts along the direction of a unit vector $\textbf{n}$ and it changes randomly with time as follows
\begin{equation}
    \dfrac{d\textbf{n}}{dt} = \mathbf{\zeta}(t) \times \textbf{n},
    \label{eq:direction}
\end{equation}
\noindent here, $\mathbf{\zeta}(t)$ is also a Gaussian white-noise random vector, having moments

\begin{equation}
	\left<\mathbf{\zeta}(t)\right>=0, \hspace{5mm}
	\left<\mathbf{\zeta}_{\alpha}(t^{\prime})\mathbf{\zeta}_{\beta}(t^{\prime\prime})\right>=2 D_{r}\delta_{\alpha\beta}\delta(t^{\prime}-t^{\prime\prime})
	\label{eq:active-random-force}
\end{equation}

\noindent where $D_r$ denotes the rotational diffusion coefficient. The translational ($D_t$) and rotational diffusion ($D_r$) coefficients of a spherical particle of size $\sigma$ are related via $D_r = \dfrac{3D_t}{\sigma^2}$. The polar form of the stochastic Eq.~\ref{eq:direction} can be obtained by transforming the Cartesian coordinates into the form of spherical coordinates $(\text{cos}\phi \text{sin}\theta, \text{sin}\phi \text{sin}\theta, \text{cos}\theta)$;

\begin{equation}
    \dfrac{\text{d}\theta}{\text{d}t} = \zeta_y \text{cos}\phi - \zeta_x \text{sin}\phi
    \label{eq:theta}
\end{equation}

\begin{equation}
    \dfrac{\text{d}\phi}{\text{d}t} = \zeta_z - \zeta_x \dfrac{\text{cos}\theta}{\text{sin}\theta} \text{cos}\phi - \zeta_y \dfrac{\text{cos}\theta}{\text{sin}\theta} \text{sin}\phi
    \label{eq:phi}
\end{equation}

\noindent where $\zeta_x$, $\zeta_y$, and $\zeta_z$ are the Cartesian coordinates of the Gaussian white noise $\zeta$. We can alternatively express the strength of active force $\textbf{F}_\text{a}$ in terms of a dimensionless quantity, P\`{e}clet number Pe, defined as $\text{Pe} = \frac{\text{F}_{\text{a}} \sigma}{k_\text{B} \text{T}}$. Therefore, $\text{Pe} = 0$ corresponds to the passive case and for the passive particles of polymer network. Thus, the self-propelled particles in our simulations are active Brownian particles (ABPs), having non-Gaussian intermediate time dynamics, even in the free space and therefore, differs from the model like active Ornstein-Uhlenbeck particle (AOUP), which is Gaussian by construction.\cite{chaki2019effects,jeon2022active,dabelow2021irreversibility,chaki2018entropy,caprini2022parental,das2018confined,theeyancheri2022silico} \\ \\
\noindent All the simulations were carried out in LAMMPS, a freely available open-source molecular dynamics simulation package.\cite{plimpton1995fast} The integration time step $\Delta t = 5 \times 10^{-4} \tau$ is chosen to be a constant in all the simulations. All the simulations are performed using the Langevin thermostat and the equation of motion (Eq.~\ref{eq:langevineq3}) integrated using the velocity Verlet algorithm in each time step. First, we equilibriate the system by running the simulations sufficiently long, such that  the average monomer-monomer distance (mesh size) is nearly constant in the range $\sigma_\text{mesh} = 0.95-1.1$, this also gives a rough measure of the mesh size for the polymer gel (Fig.~\ref{fig:S1_PDF_Mesh_size}). Thereafter, all the production simulations are carried out for 5 $\times 10^8$ steps. \\ \\
\noindent In our simulations, $\sigma$, $\epsilon$, and $m$ are taken as the fundamental units of length, energy, and mass respectively. Thus, the unit of time is $\tau=\sigma \sqrt{\frac{m}{\epsilon}}$. All other physical quantities are therefore reduced accordingly, expressed in terms of these fundamental units, $\sigma, \epsilon$, and $m$, and presented in dimensionless forms.

\section{Results and discussion} \label{sec:Results}
\noindent In order to study the influence of the self-propulsion on the dynamics of the tracer particles inside the polymer network, we compute the time-and-ensemble average of mean-square displacement (MSD) $\left(\left<\overline{\Delta r^{2}(\tau)}\right>\right)$ as a function of lag time $\tau$. In the first step, we quantify the time-averaged MSD from a single trajectory \textit{via}:

\begin{equation}
    \overline{\Delta r^{2}(\tau)} = \frac{1}{\text{T}_\text{max}-\tau} \int_{0}^{\text{T}_\text{max}-\tau} {\left[ \textbf{r}(t+\tau) - \textbf{r}(t)\right]}^2  dt,
\end{equation}

\noindent where $\text{T}_\text{max}$ is the total simulation run time, $\textbf{r}(t)$ represents the position of the tracer at initial time t, and $\textbf{r}(t+\tau)$ denotes that after lag time $\tau$. This time-averaged MSD based on a single trajectory becomes unreliable, in particular, as the lag time $\tau$ becomes long. Therefore, we obtain the time–ensemble–average MSD by performing double averaging, in other words ensemble-averaging over the time-averaging to get smoother MSD profiles,

\begin{equation}
    \left<\overline{\Delta r^{2}(\tau)}\right> = \frac{1}{N} \sum_{i=1}^{N}{\overline{\Delta r_{i}^{2}(\tau)}},
\end{equation}

\noindent where $N$ is the total number of independent trajectories. We run two independent simulations each for 25 tracer particles for a given set of parameters, so in our simulations, $N = 50$. Initially, we simulate the passive ($\text{F}_\text{a}$ = 0) and active ($\text{F}_\text{a} \neq 0$) tracer particle in a free space and compute $\left<\overline{\Delta r^{2}(\tau)}\right>$
as a function of lag time $\tau$ to validate the set of parameters used for our simulations. We fit the numerically calculated $\left<\overline{\Delta r^{2}(\tau)}\right>$ curves with the following analytical expression for an ABP (Fig.~\ref{fig:S2_MSD_ABP_Free}).\cite{du2019study}

\begin{equation}
    \left<{\Delta r^2(\tau)}\right> = 6D_{t}\tau + \frac{2\text{F}_\text{a}^2 \tau_\text{R}}{\gamma^2} \left[ \tau + \tau_\text{R} \left(e^{-\frac{\tau}{\tau_\text{R}}} - 1\right) \right]
    \label{eq:Analytical}
\end{equation}

\noindent Here $D_{t}$ is the thermal translational diffusion coefficient, $\text{F}_\text{a}$ is magnitude of the self-propulsion force and $\tau_\text{R}$ is the persistence time defined as $\tau_\text{R} = \dfrac{1}{2D_{r}}$, where $D_{r}$ is the thermal rotational diffusion coefficient. Therefore, one expects $\left<{\Delta r^2(\tau)}\right> \approx 6 D_t \tau$  for $\tau < \tau_\text{R}$, $\left<{\Delta r^2(\tau)}\right> =  6 D_t \tau + \left( \frac{2\text{F}_\text{a}^2}{e \gamma^2}\right)\tau^2$ for $\tau \simeq \tau_\text{R}$, and $\left<\Delta r^2(\tau)\right> = (6D_{t} + \frac{2\text{F}_\text{a}^2 \tau_\text{R}}{\gamma^2})\tau$ for $\tau > \tau_{R}$ (Fig.~\ref{fig:S2_MSD_ABP_Free}). \\

\begin{figure*}[ht]
    \centering
    \includegraphics[width=1.0\linewidth]{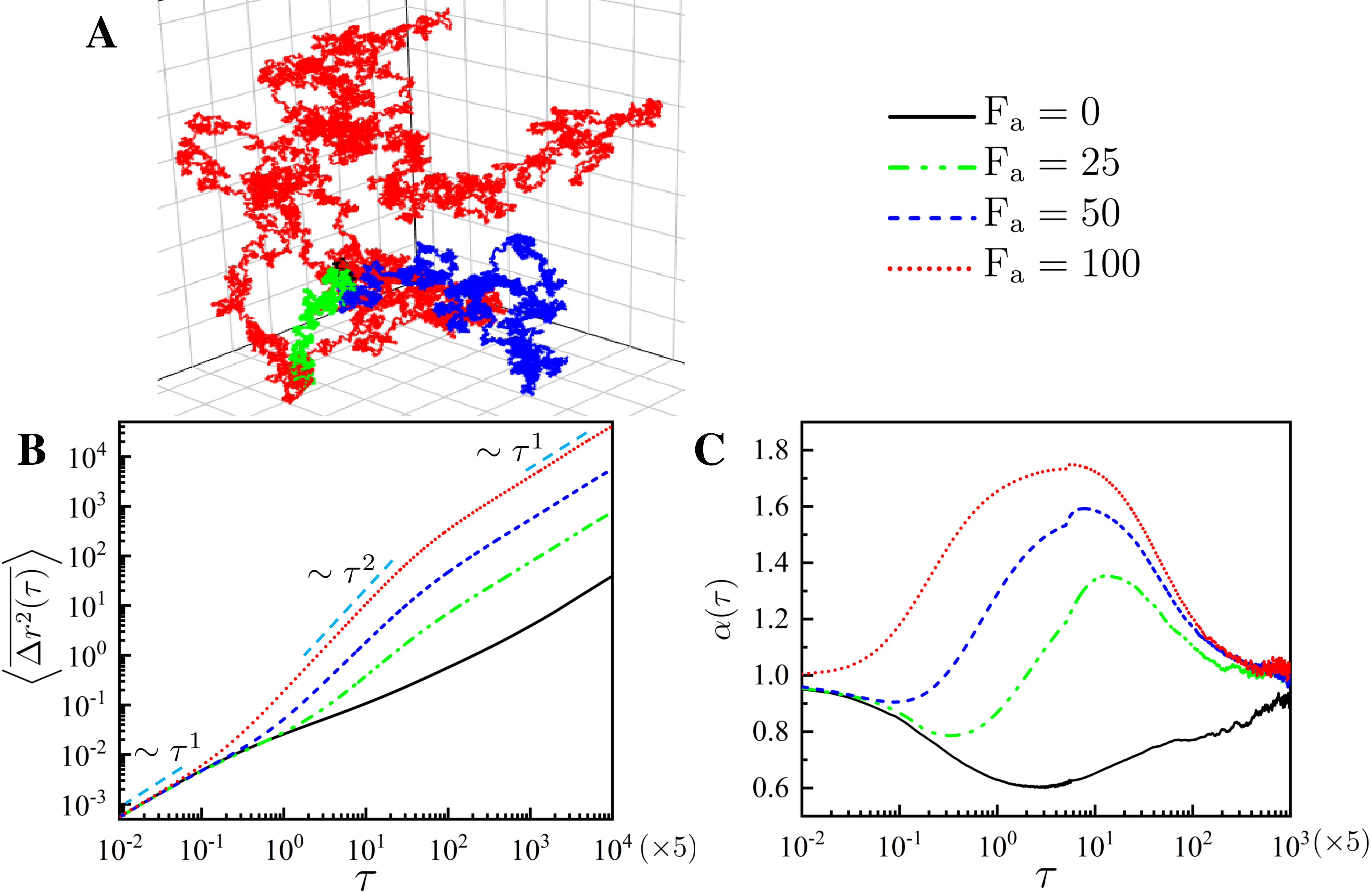}
  \caption{(A) The single trajectory plot, log-log plot of (B) $\left<\overline{\Delta r^{2}(\tau)}\right>$ $vs$ $\tau$, log-linear plot of (C) $\alpha(\tau)$ of the attractive ($\epsilon = 2.0$) self-driven tracer particle with the monomers of gel for different $\text{F} _\text{a}$ at fixed $\sigma = 1.0$ and k = 10.}
    \label{fig:traj_MSD_alpha}
\end{figure*}

\noindent Thereafter, to investigate the effect of crowding on the dynamics of the tracer particle we compute $\left<\overline{\Delta r^{2}(\tau)}\right>$ and time exponent $\alpha(\tau)$ defined as $\alpha(\tau) = \dfrac{\text{d} \log \left(\left<\overline{\Delta r^{2}(\tau)}\right>\right)}{\text{d} \log (\tau)}$. First, we consider the following set of parameters, $\sigma = 1.0$ (comparable to the mesh size) and k = 10 with varying $\text{F}_\text{a}$. We observe that the self-propulsion always leads to faster dynamics with increasing $\text{F} _\text{a}$, as evident from the plot of trajectories in three-dimension shown in Fig.~\ref{fig:traj_MSD_alpha}A. The plots of $\left<\overline{\Delta r^{2}(\tau)}\right>$ against $\tau$ in log-log scale and corresponding time exponents $\alpha(\tau)$ in log-linear scale shown in Fig.~\ref{fig:traj_MSD_alpha}(B, C), respectively. For a very short time, the dynamics is almost normally diffusive ($\alpha(\tau) \approx 1.0$). The motion of tracer crosses over from the diffusive to subdiffusive ($\alpha(\tau) < 1$) in slightly longer time scales when the active force is not high enough ($\text{F} _\text{a}$ = 100). The subdiffusive behaviour is more pronounced for the passive case ($\text{F} _\text{a} = 0$) at the intermediate time. It is due to the confined motion inside the polymer mesh created by the surrounded monomers of the gel (Movie S1). Whereas, the active tracer makes a transition from subdiffusive to superdiffusive ($\alpha(\tau) > 1$) motion from short to moderate time scale. This happens since the self-propulsion always helps the tracer to move from one mesh to another mesh and the tracer performs a persistent motion for a while, leading to superdiffusion ($\alpha(\tau) > 1$) (Movie S2). There is competition between the confinement in the the polymer network, leading to subdiffusion, and the persistent motion of the active tracer, resulting superdiffusive dynamics. For very high activity ($\text{F} _\text{a} = 100$), the dynamics always transform from diffusive to superdiffusive in short to intermediate time scales, which indicates that the activity dominates over the cage effect. $\left<\overline{\Delta r^{2}(\tau)}\right>$ of the self-propelled tracer particle grows faster with $\text{F} _\text{a}$, as demonstrated in Fig.~\ref{fig:traj_MSD_alpha}B. In the long time, the direction of the self-propulsion gets randomized due to the collisions and the particle undergoes a random walk that leads to a diffusive ($\alpha \approx 1$) motion of the tracer and $\left<\overline{\Delta r^{2}(\tau)}\right>$ becomes linear in time with an enhanced diffusion coefficient. \\

\begin{figure*}[ht]
    \centering
    \includegraphics[width=1.0\linewidth]{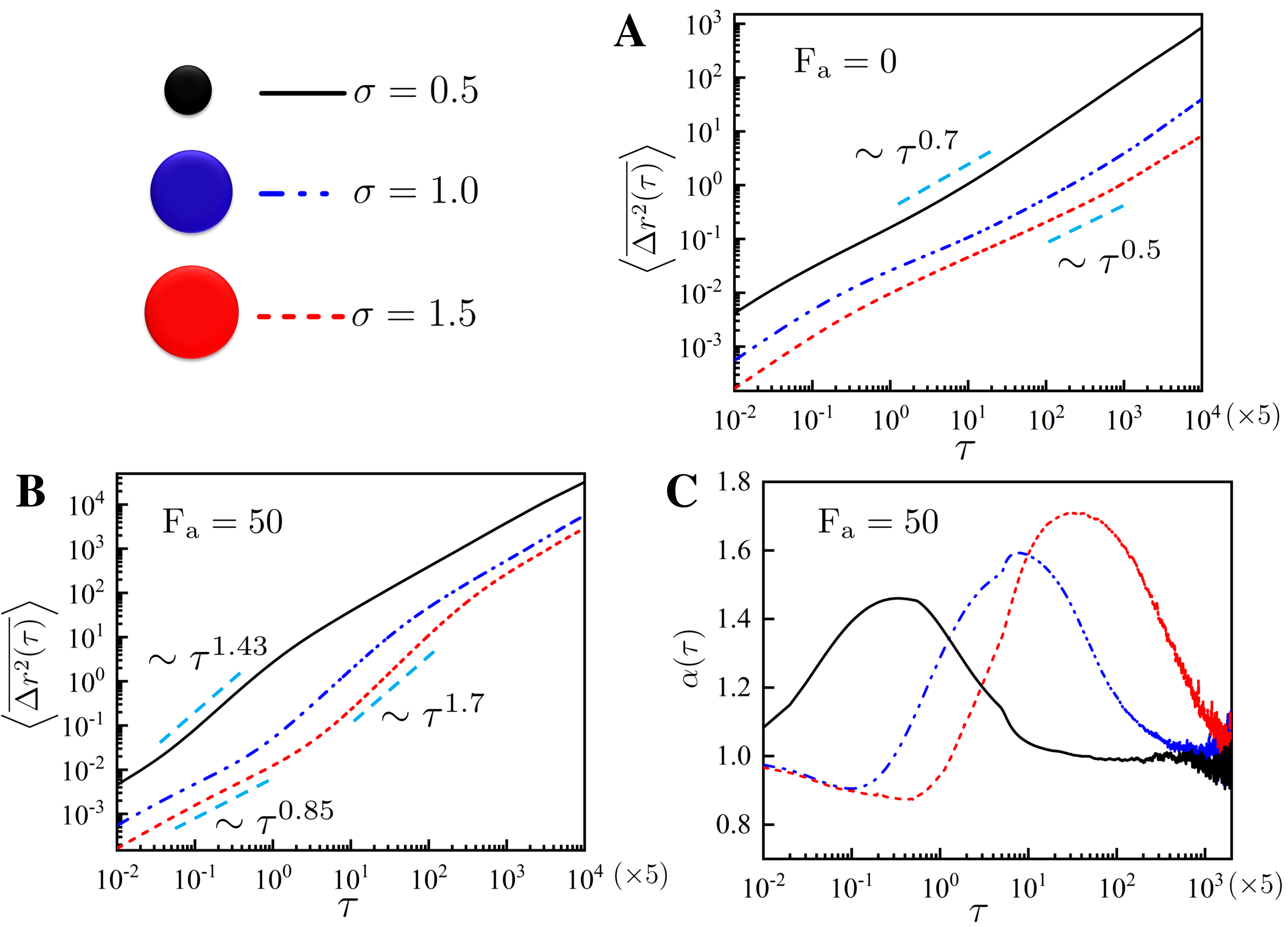}
  \caption{Log-log plot of $\left<\overline{\Delta r^{2}(\tau)}\right>$ $vs$ $\tau$ of the (A) passive $(\text{F}_\text{a} = 0)$ and (B) active ($\text{F}_\text{a} = 50$) sticky self-propelled tracer ($\epsilon = 2.0$) for different values of $\sigma = 0.5, 1.0, 1.5$ at fixed k = 10. Log-linear plot of (C) $\alpha(\tau)$ $vs$ $\tau$ of the plot (B).}
    \label{fig:MSD_size}
\end{figure*}

\noindent Further, we inspect the anomalous diffusion of the passive and active tracers in the polymer network by changing the tracer size $\sigma = 0.5 - 1.5$, keeping k 
and $\text{F} _\text{a}$ constant. For the passive case, tracer particles with the smaller size $\sigma = 0.5$ than mesh size can pass through the cages easily without being much interrupted by the obstacles and the dynamics found weakly subdiffusive ($\alpha \sim 0.7$), while tracer with the bigger size ($\sigma = 1.5$) than mesh size are transiently trapped inside the cages and $\left<\overline{\Delta r^{2}(\tau)}\right>$ shows strong subdiffusion ($\alpha \sim 0.5$) motion at the intermediate time (Fig.~\ref{fig:MSD_size}A). It is expected that the effect of confinement becomes more profound on increasing the tracer size and due to that, the motion of the tracer slows down leading to strong subdiffusive behavior (Movie S3). MSD plots in Fig.~\ref{fig:MSD_size}A display that the curve for a relatively bigger tracer always lies below that for a smaller one throughout. A similar trend is observed in Fig.~\ref{fig:MSD_size}B for an active tracer ($\text{F} _\text{a} = 50$). For this high activity, ($\text{F} _\text{a} = 50$), a tracer with the smaller size $\sigma = 0.5$ always undergoes a superdiffusive regime before achieving the normal diffusion regime like an ABP in free space. Note that in case of a relatively bigger tracer, the time exponent $\alpha(\tau)$ first decreases to a value of $\alpha(\tau) \sim 0.85$, indicating a subdiffusive behavior, then increases to the value of $\alpha(\tau) \sim 1.7$, implying a superdiffusive regime. The subdiffusion behaviour of the probe with the larger size $\sigma = 1.5$ is more pronounced than the relatively smaller probe $\sigma = 1.0$ (Fig.~\ref{fig:MSD_size}C) in short time scales. The confinement effect of the polymer network becomes stronger as the particle size increases. As time progresses, the tracer particle starts to feel the effect of the activity, and comes out from the cage, leading to the increase of $\alpha(\tau)$ (Movie S4). More interestingly, careful observation indicates that superdiffusion with the larger value of $\alpha(\tau)$ in case of bigger tracers, as depicted in Fig.~\ref{fig:MSD_size}C. As we discussed in Section~\ref{sec:Simulation}, the rotational diffusion coefficient $D_r$ scales as $\sigma^{-3}$ for an ABP and the persistence time defines by $\tau_\text{R} = {2D_r}^{-1}$, which implies that the rotation becomes harder for bigger particle and therefore, it moves more persistently along a direction as the tracer gets larger size.\cite{du2019study,yuan2019activity} \\

\begin{figure*}[ht]
    \centering
    \includegraphics[width=1.0\linewidth]{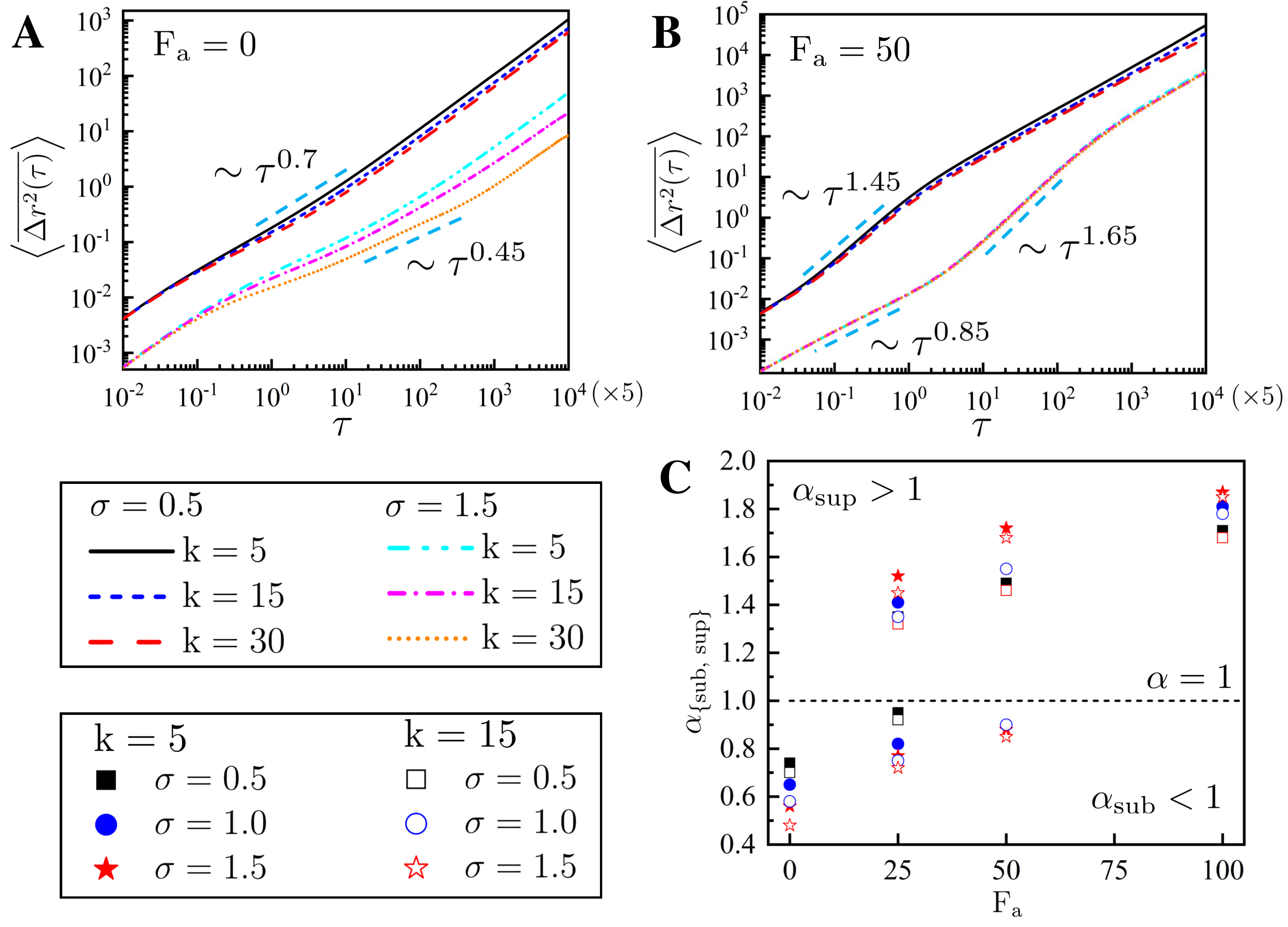}
  \caption{Log-log plots of $\left<\overline{\Delta r^{2}(\tau)}\right>$ $vs$ $\tau$ of the (A) passive $(\text{F}_\text{a} = 0)$ and (B) active ($\text{F}_\text{a} = 50$) sticky self-propelled tracer ($\epsilon = 2.0$) for different values of k = 5, 15, 30, and $\sigma = 0.5, 1.5$. Plot of (C) $\alpha_{\text{\{sub, sup\}}}$ $vs$ $\text{F} _\text{a}$: subdiffusion ($\alpha_\text{sub}$) and superdiffusion ($\alpha_\text{sup}$) exponents are represented below and above the dashed-line ($\alpha = 1$) respectively, for $\sigma = 0.5, 1.5,$ and k = 5, 15.}
    \label{fig:MSD_stiff}
\end{figure*}

\noindent In Fig.~\ref{fig:MSD_stiff}(A,B), we quantify $\left<\overline{\Delta r^{2}(\tau)}\right>$ of the passive and active tracers by changing the stiffness (k) of the polymer network for tracer with diameters $\sigma =$ 0.5 and 1.5. On increasing the value of k, the particles of the polymer network become less mobile and form nearly static meshes, which suppresses the motion of the passive tracer particles. From Fig.~\ref{fig:MSD_stiff}A we notice that the effect of k is more significant for a bigger ($\sigma = 1.5$) tracer in comparison to the smaller one ($\sigma = 0.5$) since a particle with a smaller size can escape from the meshes easily, while the bigger particles are tightly placed inside the meshes and transiently trapped. It is noticeable that at a very high value of k, the exponent $\alpha$ approaches a value of 0.45. On the other hand, self-propulsion always promotes the motion of active tracer to explore larger volume that too persistently inside the polymer network. Therefore, $\left<\overline{\Delta r^{2}(\tau)}\right>$ of the self-propelled tracers grow faster with $\text{F} _\text{a}$, as demonstrated in Fig.~\ref{fig:MSD_stiff}B. Interestingly, for higher activity ($\text{F} _\text{a}$ = 50), the self-propulsion controls the dynamics of the tracer and washes out the effect of network stiffness, leading to the overlapping of $\left<\overline{\Delta r^{2}(\tau)}\right>$ corresponding to different values of k (Fig.~\ref{fig:MSD_stiff}B). Consequently, to measure the degree of subdiffusion and superdiffusion, we calculate the anomalous diffusion exponent $\alpha_{\text{\{sub, sup\}}}$ as a function of $\text{F} _\text{a}$, as presented in Fig.~\ref{fig:MSD_stiff}C, where $\alpha_\text{sub}$ and $\alpha_\text{sup}$ are the minimum and maximum values of $\alpha(\tau)$ for subdiffusion ($\alpha(\tau) < 1$) and superdiffusion ($\alpha(\tau) > 1$) dynamics, respectively. In Fig.~\ref{fig:MSD_stiff}C, we see that $\alpha_{\text{\{sub, sup\}}}$ increase with $\text{F} _\text{a}$ since self-propulsion reduces the subdiffusion ($\alpha(\tau)$ increases) and enhances the superdiffusion. For bigger tracers, initially one sees strong subdiffusion, owing to its tight motion inside the mesh. However, eventually self-propulsion takes over as the time progresses and dynamics becomes strongly superdiffusive. This is essentially due to the fact that the persistence time scales linearly with the volume of the tracer. We also report that both $\alpha_\text{sub}$ and $\alpha_\text{sup}$ decrease on increasing k. One can clearly see that the differences between the values of $\alpha_\text{sub}$ and $\alpha_\text{sup}$, corresponding to different values of k are more profound for passive and weakly active tracers. On the other hand, values are extremely close for highly active tracers (large $\text{F} _\text{a}$). Thus, for the highly active tracers, the stiffness of the network does not control the dynamics, rather the activity does.

\noindent Role of self-propulsion in assisting the escape of the tracer particles from the traps, created by the polymer network becomes visible in case of highly sticky tracer ($\epsilon=4$). If the tracer is passive and has strong affinity towards the polymer beads ($\epsilon=4$), a strong subdiffusion is observed with a lowest possible value $\alpha(\tau)=0.5$ (Fig.~\ref{fig:S3_MSD_WCA_E2E4}A). While if the tracer is self-propelled ($\text{F}_\text{a} = 50$), the value of minimum subdiffusive $\alpha(\tau)$ increases to $0.77$ and a strong superdiffusion with $\alpha=1.52$ at later time (Fig.~\ref{fig:S3_MSD_WCA_E2E4}B). On the other hand, for the non-sticky tracer interacting with the network monomers via repulsive WCA potential the dynamics is weakly subdiffusive (Fig.~\ref{fig:S3_MSD_WCA_E2E4}A) for the passive tracer and for the active tracer it is highly superdiffusive (Fig.~\ref{fig:S3_MSD_WCA_E2E4}B). \\

\begin{figure*}[ht]
    \centering
    \includegraphics[width=1.0\linewidth]{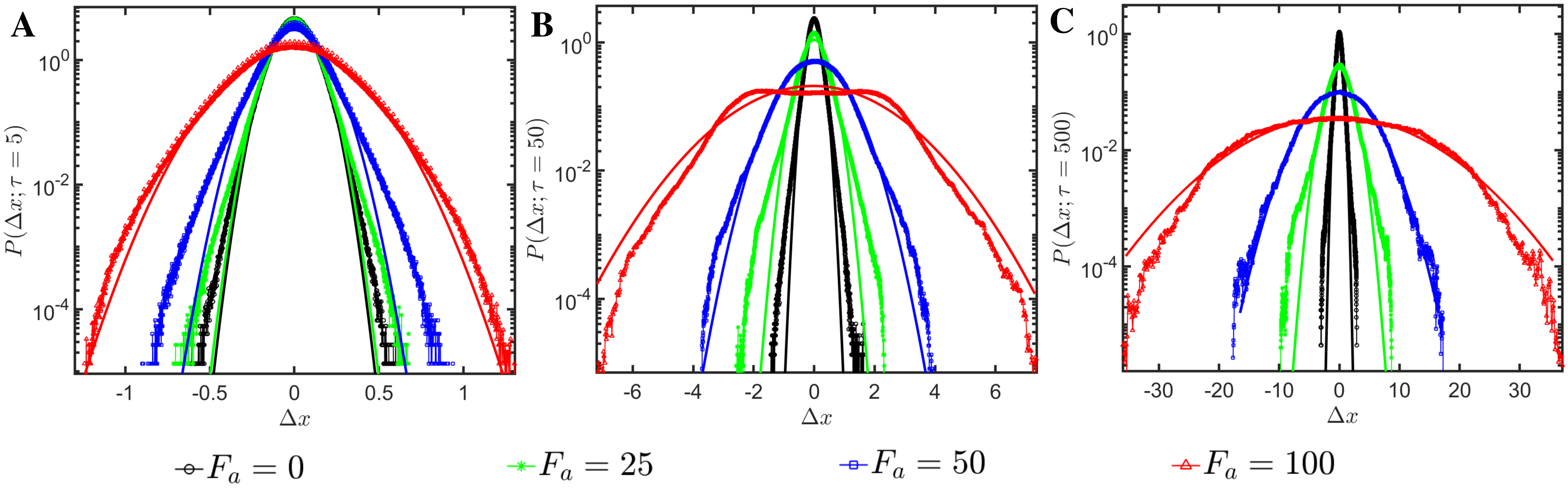}
    \caption{Plots of $\text{P}(\Delta x; \tau)$ for the sticky ($\epsilon = 2.0$) tracer particle as a function of $\text{F} _\text{a}$ at different lag times (A) $\tau = 5$, (B) $\tau = 50$, and (C) $\tau = 500$ for $\sigma = 1.0$ and k = 10. Solid lines represent the Gaussian fittings.}
    \label{fig:PDF_active}
\end{figure*}

\begin{figure*}[ht]
    \centering
    \includegraphics[width=1.0\linewidth]{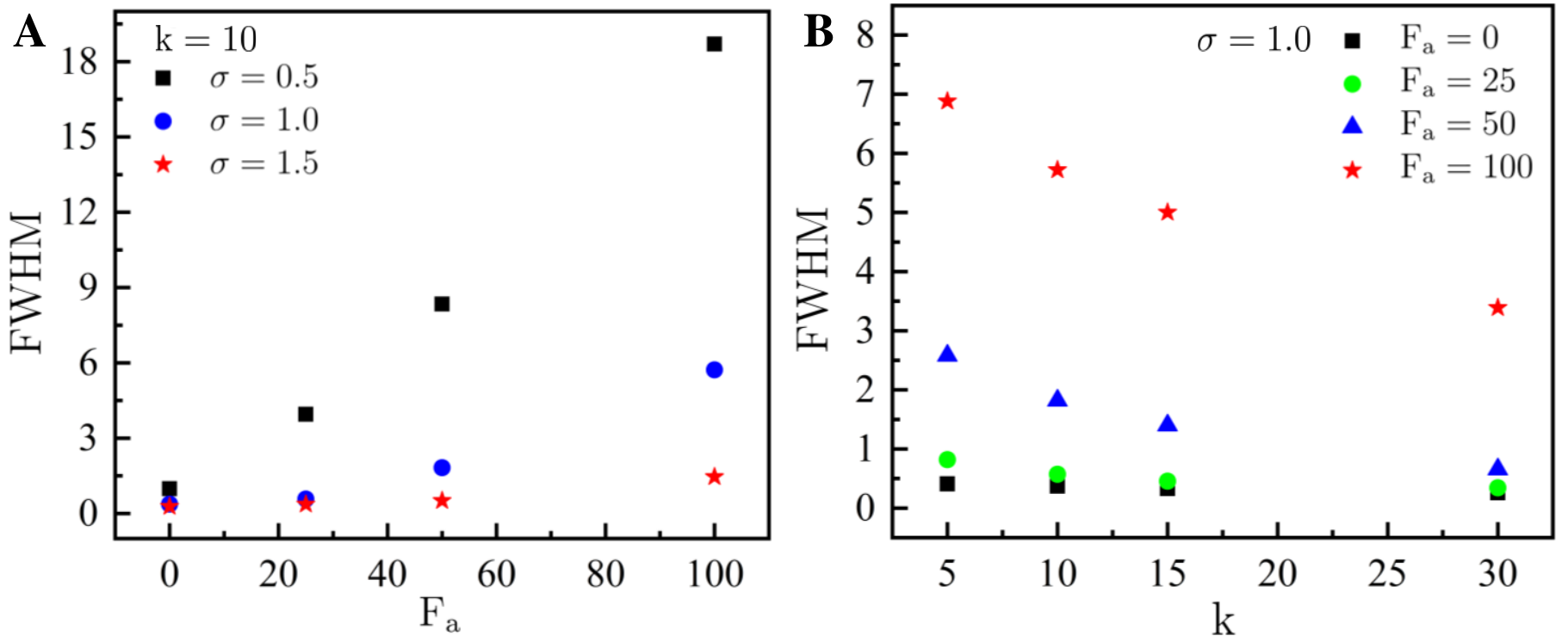}
    \caption{Plots of (A) full width half maxima (FWHM) $vs$ $\text{F}_\text{a}$ as a function of $\sigma$ for k = 10 and (B) FWHM $vs$ k as a function of $\text{F}_\text{a}$ for $\sigma = 1.0$ at the intermediate lag time $\tau = 50$.}
    \label{fig:FWHM_size_stiff}
\end{figure*}

\noindent In typical single particle tracking experiments, one way to quantify the dynamics of the tracer is to construct of the tracer's van-Hove correlation or the displacement distribution from its recorded trajectories.\cite{wagner2017rheological,anderson2021subtle,rose2020particle,levin2021measurements} From our simulations, we also compute the self part of the van-Hove correlation function (displacement probability distribution function). We compute it for one dimension (computing the same for three dimension is straightforward), which is defined as follows, $P(\Delta x; \tau) \equiv \left\langle \delta(\Delta x - (x(t+\tau)-x(t)))\right\rangle$, where $x(t+\tau)$ and $x(t)$ are the positions of the tracer along the $x$ direction at time $(t+\tau)$ and t, respectively. Here, $P(\Delta x; \tau)$ is essentially the histogram obtained from several independent trajectories. Thus, the distributions of the displacements are time-and-ensemble averaged. In Fig.~\ref{fig:PDF_active}, we report $P(\Delta x; \tau)$ of the tracer with $\sigma = $1.0 inside the network having k = 10 for different values of $\text{F} _\text{a}$. We also depict trial fittings of these van-Hove distributions with a Gaussian distribution, $P_g (\Delta x; \tau)=\frac{1}{\sqrt{2\pi\left<\Delta x^2\right>}}\exp\left({-\frac{\Delta x^2}{2\left<\Delta x^2\right>}}\right)$. Therefore, any deviation of the van-Hove distributions from the Gaussian curves indicates non-Gaussianity. Please note that here $\left<\Delta x^2\right>$ in $P_g (\Delta x; \tau)$ is also a fitting parameter and does not necessarily represent the actual variance for the tracer concerned. At short time ($\tau$ = 5) for the passive and weakly active tracer ($\text{F} _\text{a} = 25$), the distribution curves overlap but for higher values of $\text{F} _\text{a} = 50, 500$, $P(\Delta x; \tau)$s become broader, which accounts for the larger displacement of the tracer inside the polymer network (Fig.~\ref{fig:PDF_active}A). As the time progresses, $P(\Delta x; \tau)$ broaden further with increasing $\text{F}_\text{a}$, as displayed in Fig.~\ref{fig:PDF_active}(B-C). These van-Hove are not only broad but also flat over a length scale $\sim -3\sigma$ to $+3\sigma$ for $\text{F} _\text{a} = 100$ at $\tau = 50$ and $-25\sigma$ to $+25\sigma$ for $\tau = 500$. This indicates that at higher self-propulsion the tracer moves between the meshes without getting trapped. In other words, shorter and larger displacements become equally probable. In addition, it can also be seen from the plots that the deviation from Gaussianity is more pronounced at the intermediate time. For very short or long time, the deviations are less. This is essentially due the local heterogeneity of the medium which washes out in the long time for passive or weakly active tracers. However, for a strongly active tracer (say $\text{F} _\text{a} = 100$) van-Hove distribution is broad and flat, even in the long time limit. Flattening of the van-Hove distribution in this case indicates that at such higher self-propulsion, the tracer moves like a free persistent walker over a wide length scale. We also evaluate the full-width half maxima (FWHM) of van-Hove distributions to examine the effects of tracer size $\sigma$ and stiffness k of the polymer network on the $P(\Delta x; \tau)$ for a range of active forces, as illustrated in Fig.~\ref{fig:FWHM_size_stiff}. Higher the activity, broaden the curves are. FWHM always increases with increasing $\text{F} _\text{a}$. On the other hand, with increasing k and $\sigma$, $P(\Delta x; \tau)$ becomes narrower, reflecting confined motion. Apart from this, the non-sticky tracers move faster than the sticky ones due to the repulsive interactions between the tracer and the gel particles, which results broader displacement distributions. On the other hand, $P(\Delta x; \tau)$ becomes narrower with increase in the binding affinity of the tracer towards the polymer network (Fig.~\ref{fig:S4_PDF_WCA_E2E4}A). In the case of active tracer ($\text{F} _\text{a} = 50$), the distribution curves are relatively broader and the differences are small (Fig.~\ref{fig:S4_PDF_WCA_E2E4}B). This indicates that the effect of stickiness becomes insignificant on increasing the self-propulsion. \\

\section{Conclusions} \label{sec:Conclusions}
\noindent Motivated by the phenomena of transport of biomolecules, bacteria and synthetic nanomotors through mesh-like environments, such as mucus membrane, nuclear pore complex, porous media, \cite{pederson2000diffusional,amblard1996subdiffusion,bhattacharjee2019bacterial,sigalov2010protein,sumino2012large,loose2014bacterial,woolley2003motility,wu2021mechanisms} we use computer simulations to elucidate the mechanism of transport of active tracers through a model polymer network, created on a diamond lattice. Our results show that the dynamics of tracer particle always gets enhanced on switching on and increasing the self-propulsion force or activity. We have found that the dynamics of the passive tracer is always subdiffusive at an intermediate time, whereas the self-propelled tracer exhibits a transition from a short-time subdiffusion to an intermediate-time superdiffusion. In the absence of activity, the particle is confined inside the mesh, especially when comparable to or bigger than the mesh size. This leads to subdiffusion for a short time, but due to the self-propulsion, the particle escapes from its caged motion and exhibits superdiffusion in the relatively longer time.  This transition between subdiffusion and superdiffusion thus can be attributed to the competition between the confined motion of the tracer due to its size and sticky interaction with the mesh particles and the self-propulsion it possesses. The persistence motion of the self-propelled particle increases with its size, as the persistence time for an active Brownian particle scales linearly with its volume. Therefore, a bigger tracer has a higher value of $\alpha(\tau)$ than the smaller one.  Additionally, we found that the effect of the polymer stiffness is more profound in the case of a bigger particle as the smaller one can travel between the meshes without feeling the network. But on increasing the stiffness of the network, the dynamics of the monomers of the network slows down and suppresses the motion of tracers, comparable to or bigger than the average mesh size. This leads to a substantial decline in the value of $\alpha(\tau)$. At very high activity, \textit{e.g.} $\text{F}_\text{a} = 100$, the persistent motion dominates, short-time subdiffusion disappears and the particle switches over gradually from diffusive to superdiffusive motion. To achieve a deeper understanding of the transport process, we also investigated the van-Hove correlation function and found that the distribution increasingly becomes broader with increasing $\text{F}_\text{a}$, implying a larger displacement of the active tracer with increasing activity. Especially, in the case of higher activity ($\text{F}_\text{a} = 100$), displacement distribution displays a broad and flat regime at an intermediate time, which might be the consequence of the larger persistence length of the tracer. Surprisingly, the displacement distributions deviate significantly from Gaussianity at higher self-propulsion, even at relatively longer time, indicating unrestricted transport of the tracer inside the network.  In a future work, we plan to extend our current study to include asymmetrical self-propelled particles. We expect that our findings can be useful in understanding the mechanism of transport of active agents, such as bacteria, self-powered nano/micro devices in mesh-like complex environments. \\

\begin{acknowledgements}
\noindent P. K. thanks the UGC for fellowship. R. C. acknowledges the SERB for funding (Project No. MTR/2020/000230 under MATRICS scheme) and IRCC-IIT Bombay (Project No. RD/0518-IRCCAW0-001) for funding. The authors thank Ligesh Theeyancheri for the helpful discussions. We acknowledge the SpaceTime-2 supercomputing facility at IIT Bombay for the computing time. The authors thank Sanaa Sharma for reading the manuscript. \\
\end{acknowledgements} 

\noindent \textbf{DATA AVAILABILITY} \\

\noindent The data that supports the findings of this study are available within the article [and its supplementary material]. \\

\renewcommand{\thefigure}{S\arabic{figure}}
\setcounter{figure}{0}
\setcounter{equation}{0}
\renewcommand{\theequation}{\Roman{equation}}
\appendix

\begin{figure*}[ht]
    \includegraphics[width=0.95\linewidth]{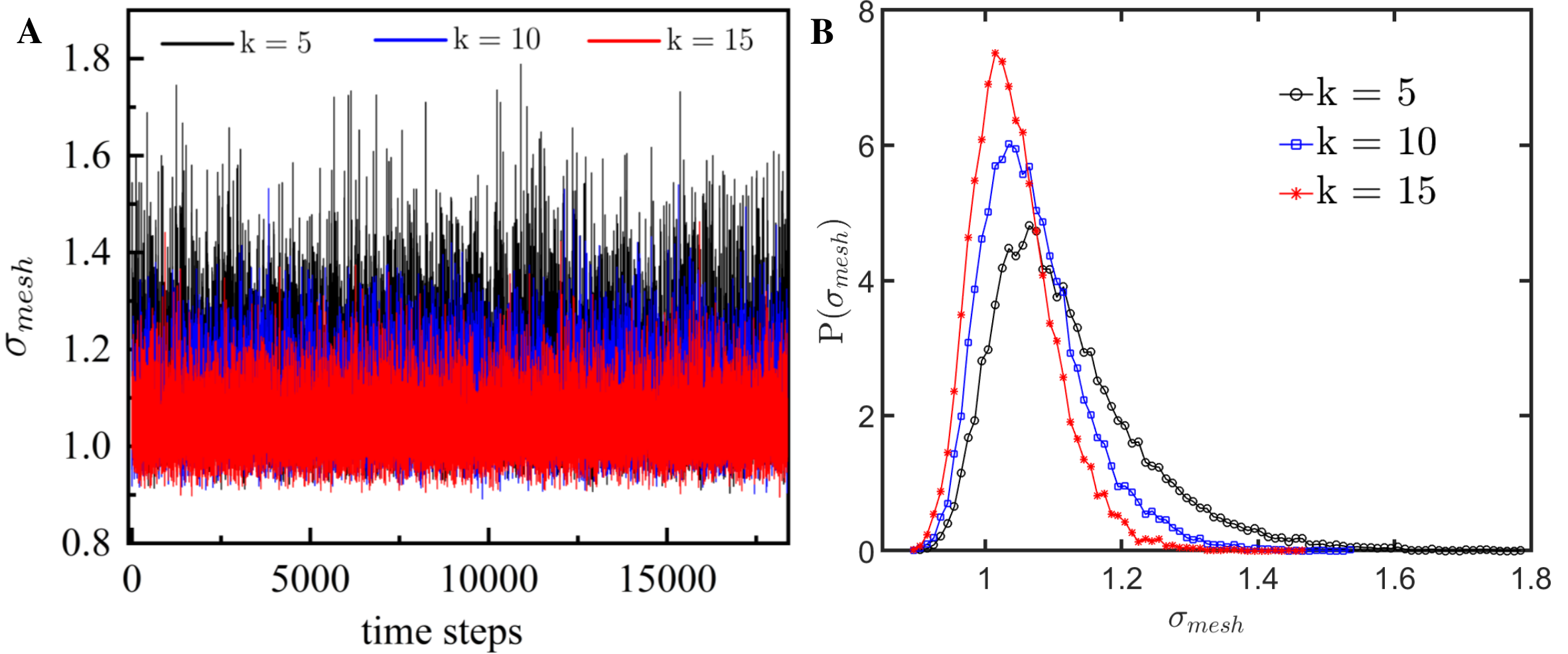}
    \caption{Plots of (A) the mesh size fluctuation ($\sigma_{mesh}$) \textit{vs} time steps and (B) mesh size distribution P($\sigma_{mesh}$) of the polymer network.}
    \label{fig:S1_PDF_Mesh_size}
\end{figure*}

\noindent \textbf{SUPPLEMENTARY MATERIAL} \\ \\
\noindent To characterize the polymer network which is constructed on a diamond lattice, we measure the average mesh size of the polymer network by considering distances between the nearest monomers of the network as shown in Fig.~\ref{fig:S1_PDF_Mesh_size}. \\

\begin{figure*}[ht]
    \includegraphics[width=0.95\linewidth]{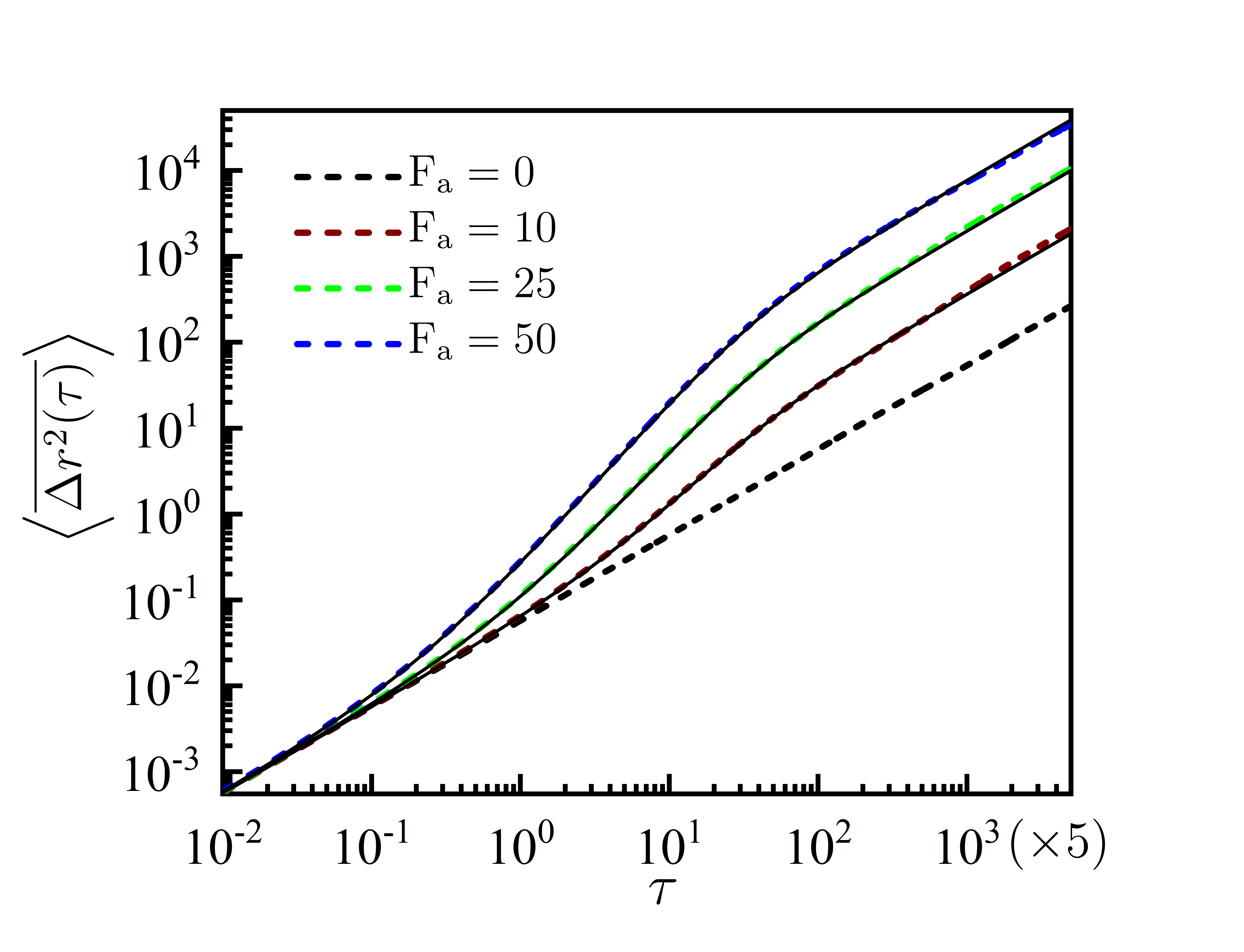}
    \caption{Log-log plot of (A) $\left<\overline{\Delta r_\text{c}^{2}(\tau)}\right>$ fitted with Eq.~(\ref{eq:Analytical_1}) (solid black lines) for the self-propelled tracer particle (ABP) in free space for various $\text{F}_\text{a}$.}
    \label{fig:S2_MSD_ABP_Free}
\end{figure*}

\noindent For method and parameter validation, we carried out the simulations of an active Brownian particle (ABP) in free space. The time-averaged mean-squared displacement $\left(\left<\overline{\Delta r^{2}(\tau)}\right>\right)$ is calculated. From the plot (Fig.~\ref{fig:S2_MSD_ABP_Free}) for $\text{F}_\text{a}$ = 0, first we have computed the thermal translational diffusion coefficient, $D_{t} = 9.35\times 10^{-5}$ and then rotational diffusion coefficient, $D_R = 2.8 \times 10^{-4}$ by using the expression, $D_r = \dfrac{3D_t}{\sigma^2}$, where $\sigma$ is the size of an ABP. Thus, the persistence time is, $\tau_{R} = \frac{1}{2 D_{r}} = 1.8 \times 10^{3}$. Using the values of $D_t$ and $\tau_R$, $\left<\overline{\Delta r^{2}(\tau)}\right>$ is fitted with the analytical expression for ABP.
\begin{equation}
    \left<{\Delta r^2(\tau)}\right> = 6D_{t}\tau + \frac{2\text{F}_\text{a}^2 \tau_{R}}{\gamma^2} \left[ \tau + \tau_{R} \left(e^{-\frac{\tau}{\tau_R}} - 1\right) \right]
    \label{eq:Analytical_1}
\end{equation}

\noindent For the passive case ($\text{F}_\text{a}$ = 0), $\left<\overline{\Delta r^{2}(\tau)}\right>$ is always diffusive $\left(\left<\overline{\Delta r^{2}(\tau)}\right> \sim \tau\right)$ with the diffusion coefficient $D_{t}$. In case of self-propelled tracer, $\left<\overline{\Delta r^{2}(\tau)}\right>$ exhibits three distinct regions: diffusive at short time ($\tau < \tau_{R}$), superdiffusive region at intermediate time ($\tau \simeq \tau_{R}$) which scales as $\left<\overline{\Delta r^{2}(\tau)}\right>$ $\sim \tau^2$, followed by enhanced diffusive region at longer time, \textit{i.e.} $\tau > \tau_{R}$ and the expression becomes $\left<\Delta r^2(\tau)\right> = (6D_{t} + \frac{2\text{F}_\text{a}^2 \tau_{R}}{\gamma^2})\tau$. $\left<\overline{\Delta r^{2}(\tau)}\right>$ grows faster with $\text{F}_\text{a}$ in comparison to the passive tracer (shown in Fig.~\ref{fig:S2_MSD_ABP_Free}). \\ \\

\begin{figure*}[ht]
    \includegraphics[width=1.0\linewidth]{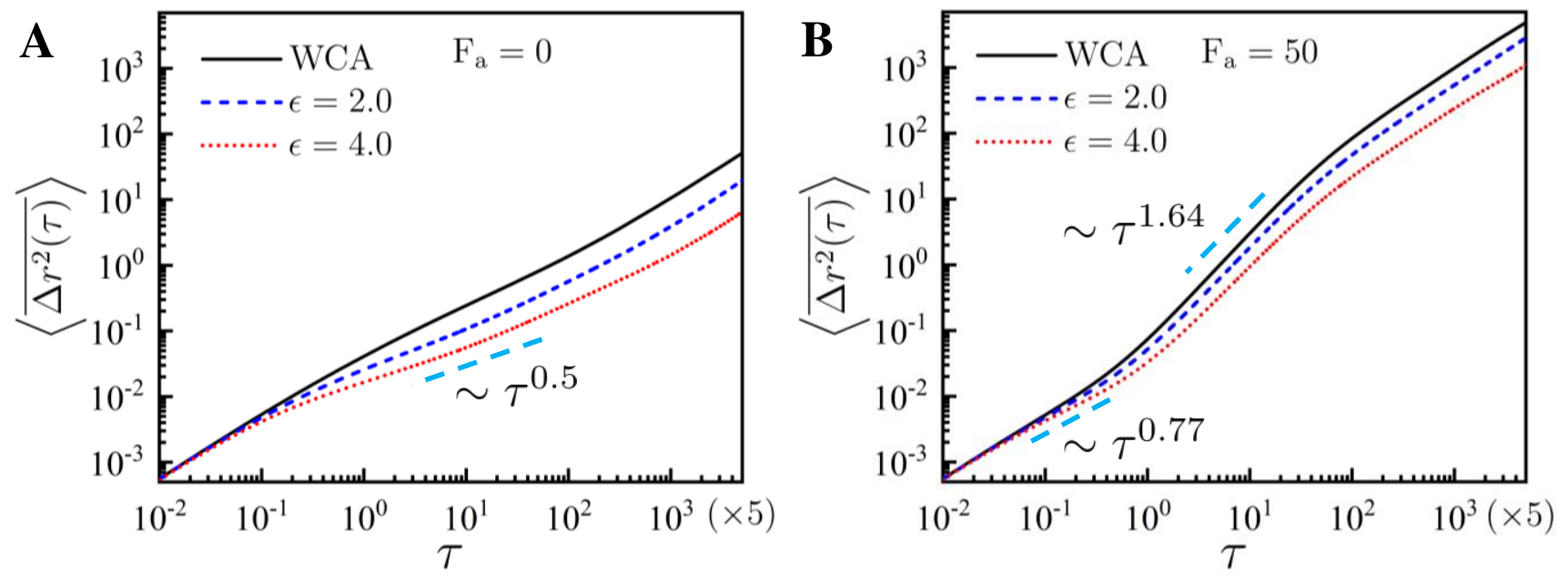}
    \caption{Log-log plots of $\left<\overline{\Delta r^{2}(\tau)}\right>$ $vs$ $\tau$ of the (A) passive ($\text{F}_\text{a} = 0$) and (B) active ($\text{F}_\text{a} = 50$) tracers inside the polymer network by varying the interaction of the tracers with the monomer of the network at $\sigma = 1.0$ and $k = 10$.}
    \label{fig:S3_MSD_WCA_E2E4}
\end{figure*}

\begin{figure*}[ht]
    \includegraphics[width=1.0\linewidth]{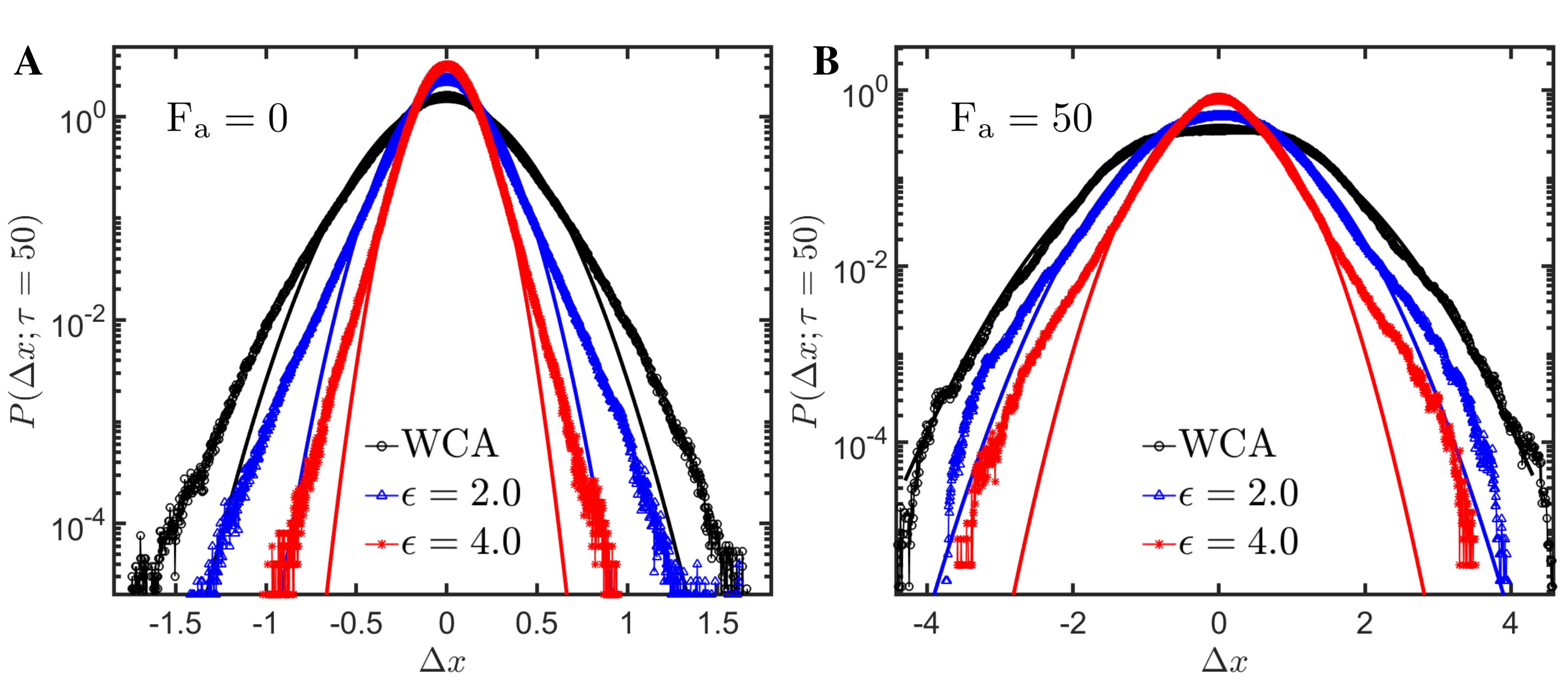}
    \caption{$P(\Delta x; \tau)$ of the (A) passive ($\text{F}_\text{a} = 0$) and (B) active ($\text{F}_\text{a} = 50$) tracers inside the polymer network for different interaction of the tracer particles with network at $\sigma = 1.0$, $k = 10$, and lag time $\tau = 50$. Solid lines represent the best Gaussian fits.}
    \label{fig:S4_PDF_WCA_E2E4}
\end{figure*}

\noindent \textbf{Movie Description} \\ 
\noindent \textbf{Movie S1:} Molecular dynamics simulation of the passive ($\text{F}_\text{a}$ = 0) tracer particles (red in color) and it is clear from the movie passive tracers are transiently trapped inside the polymer network. \\

\noindent \textbf{Movie S2:} Molecular dynamics simulation of the self-propelled ($\text{F}_\text{a}$ = 50) tracer particles. One can see that the self-propulsion helps the tracer to escape from the polymer meshes and it covers a larger space inside the network. \\

\noindent \textbf{Movie S3:} Molecular dynamics simulation of the relatively bigger particles ($\sigma = 1.5$) for $\text{F}_\text{a}$ = 0. Here, passive particles are trapped inside the meshes formed by the polymer network. \\

\noindent \textbf{Movie S4:} Molecular dynamics simulation of the relatively bigger particles ($\sigma = 1.5$) for $\text{F}_\text{a}$ = 50. Here, the dynamics of self-propelled tracer particles becomes faster leading to escape dynamics from the meshes of the polymer network. \\

\clearpage

%

\end{document}